\documentclass[structabstract]{aa}  

\usepackage{graphicx}
\usepackage{txfonts}
\usepackage{natbib}
\usepackage{color}

\bibpunct{(}{)}{;}{a}{}{,} 
\begin{document}

   \title{The UV galaxy Luminosity Function at $z$=3-5 from the CFHT Legacy Survey Deep fields\thanks{Based on observations obtained with
       MegaPrime/MegaCam, a joint project of CFHT and CEA/DAPNIA, at
       the Canada-France-Hawaii Telescope (CFHT) which is operated by
       the National Research Council (NRC) of Canada, the Institut
       National des Sciences de l'Univers of the Centre National de la
       Recherche Scientifique (CNRS) of France, and the University of
       Hawaii. This work is based in part on data products produced at
       TERAPIX and the Canadian Astronomy Data Centre as part of the
       Canada-France-Hawaii Telescope Legacy Survey, a collaborative
       project of NRC and CNRS.}
          }

   \author{Remco F.J. van der Burg
          \inst{1}
          \and
          Hendrik Hildebrandt\inst{1}
          \and  
          Thomas Erben\inst{2}
          }

   \institute{Leiden Observatory, Leiden University, Niels Bohrweg 2, 2333CA Leiden, The Netherlands\\
                 \email{vdburg@strw.leidenuniv.nl}        
                 \and
             Argelander-Institut f\"ur Astronomie, Auf dem H\"ugel 71, 53121 Bonn, Germany\\
             }

   \date{Received ...; accepted ...}

  \abstract {} {We measure and study the evolution of the UV
      galaxy Luminosity Function (LF) at $z$=3-5 from the largest
      high-redshift survey to date, the Deep part of the CFHT Legacy
      Survey. We also give accurate estimates of the SFR density at
      these redshifts.}  {We consider $\sim$ 100\,000 Lyman-break
    galaxies at $z\approx$ 3.1, 3.8 \& 4.8 selected from very deep
    $ugriz$ images of this data set and estimate their rest-frame
    1600$\rm{\AA}$ luminosity function. Due to the large survey
    volume, cosmic variance plays a negligible role. Furthermore, we
    measure the bright end of the LF with unprecedented statistical
    accuracy. Contamination fractions from stars and low-$z$ galaxy
    interlopers are estimated from simulations. From these simulations
    the redshift distributions of the Lyman-break galaxies in the
    different samples are estimated, and those redshifts are used to
    choose bands and calculate k-corrections so that the LFs are
    compared at the same rest-frame wavelength.  To correct for
    incompleteness, we study the detection rate of simulated
    galaxies injected to the images as a function of magnitude and
    redshift. We estimate the contribution of several systematic
    effects in the analysis to test the robustness of our results.}
            {We find the bright end of the LF of our $u$-dropout sample to deviate significantly from a Schechter function. If we modify the function by a recently proposed magnification model, the fit improves. For the first time in an LBG sample, we can measure down to the density regime where magnification affects the shape of the observed LF because of the very bright and rare galaxies we are able to probe with this data set. 
             We find an increase in the normalisation, $\phi^{*}$, of
              the LF by a factor of 2.5 between $z\approx5$ and
              $z\approx3$. The faint-end slope of the LF does not evolve significantly between $z\approx5$ and
              $z\approx3$. We do not find a
              significant evolution of the characteristic magnitude in
              the studied redshift interval, possibly because of insufficient
              knowledge of the source redshift
              distribution. The SFR density is found to
                increase by a factor of $\sim$2 from $z\approx 5$ to
                $z\approx 4$. The evolution from $z\approx
                4$ to $z\approx 3$ is less eminent.}{}

   \keywords{galaxies: high-redshift --
                galaxies: luminosity function --
                galaxies: evolution
               }

   \maketitle
%

\section{Introduction}
The formation and evolution of galaxies rank among the big questions
in astronomy and still await a complete explanation. According to
current theory, the formation of dark matter haloes by gravitational
instabilities is an essential first step in the formation of galaxies
\citep{eggen62}. Stars are believed to form when gas cools at the
centres of these haloes \citep{white78}, and make up the part of the
galaxy that we can observe. A number of physical processes
strongly affect this baryonic mass assembly, like the
hydrodynamics of the gas, feedback processes by supernovae and stellar
winds, possibly magnetic fields, the role of AGN, or the effects of
galaxy-galaxy interactions and mergers. For these reasons the
modelling of galaxy formation depends on many free parameters and is
not very well constrained.

Over the past decade the high redshift universe has become accessible
observationally through the use of photometric techniques.
By detection of the spectral discontinuity due to the
  redshifted Lyman-break in a multi-wavelength filter set, large and
  clean samples of high redshift star-forming galaxies can be selected
  \citep{steidel96, steidel99, 2002ARA&A..40..579G}, with low amounts
  of contamination. These samples can be used to study several
properties of the early universe. For example, by measuring
  the correlation function of these Lyman-break Galaxies (LBGs) and
  comparing it with the correlation of dark matter, the characteristic
  mass of their haloes can be determined \citep[e.g.][]{giadick01,
  ouchi04b, 2005A&A....Hildebrandt, 2007A&A...462..865H,
  hildebrandt09a}. Hubble Space Telescope observations of LBGs are
used to study how certain morphological types evolve with time
\citep{pirzkal05}. A study of the evolution of the 
UV Luminosity Function (LF) \citep{steidel96, steidel99, st2,
  bouwens07}, which is the measure of the number of galaxies per unit
volume as a function of luminosity, is another fundamental
probe in galaxy formation and evolution, because of its close
  relation to star formation processes.

Several techniques can be used to estimate the star formation rate
(SFR) in galaxies, mostly based on the existence of massive, young
stars, indicative of recent star formation. A commonly used way to
probe the existence of massive stars is the H$\alpha$ luminosity
\citep{kennicutt83}, because H$\alpha$ photons originate from the gas
ionized by the radiation of massive stars. A second star formation
indicator is the infrared (IR) luminosity originating from dust
continuum emission \citep{kennicutt98,hirashita03}. The absorption
cross section of dust is strongly peaked in the UV, and
therefore the existence of UV emitting, i.e. massive, stars is probed
indirectly. Thirdly, the UV continuum is used as a star formation
probe, with the main advantages being that the UV-emission of the
young stellar population is $directly$ observed, unlike in H-$\alpha$
and IR studies. Furthermore, this technique can be applied from the
ground to star-forming galaxies over a wide range of redshifts.
Hence, it is still the most powerful probe of cosmological evolution
of the SFR \citep{madau96}. However, information about the initial
mass function (IMF), and particularly the extinction by dust are required to
estimate the total star formation rate.

In this paper we estimate the UV LF at redshifts $z$=3-5 from the
Canada-France-Hawaii-Telescope Legacy Survey (CFHTLS) Deep, a survey
covering 4 square degrees in four independent fields spread across the
sky. Since our samples, at different redshifts, are all extracted from
the same dataset, this gives an excellent opportunity to study a
possible evolution of the LF in this redshift
interval. Several systematic effects that need to be considered when
comparing results at different redshifts from different surveys
(e.g. source extraction, masking, PSF-modelling, etc.) can be avoided
when the different redshift samples are extracted from the same
survey. Due to the large volumes we probe with our 4 square degree
survey, 
the influence of cosmic variance on the shape of the estimated LF is negligible 
\citep{trenti08}, as
  we expect cosmic variance to affect our number counts only at the
  1-5\% level \citep{somerville04}. 
  We can study the bright end of the LF with unprecedented accuracy, as these objects are rare and we are able to measure down to very low densities. This allows us to study the effect that magnification has on the observed distribution \citep[see recent results by][]{jainlima2010}, and study a possible deviation from the commonly used Schechter function.
  Furthermore, given the depth of
the stacked images, we can probe the faint end of the luminosity
function with comparable precision as the deepest ground based surveys
have done before \citep{st2}. 

The structure of this paper is as follows: In
Sect. \ref{sec:datasamples} we describe the data set we use, the LBG
selection criteria as well as the simulations that lead to the
redshift distributions and contamination fractions. In
Sect. \ref{sec:analysis} the survey's completeness and the effective
survey volumes are estimated. In Sect. \ref{sec:results} we proceed
with the resulting estimated LFs, present the best-fitting
Schechter parameters, and show how a simple magnification model can significantly improve the quality of the fit. The UV {luminosity densities (UVLD)} and
SFR densities (SFRD) are estimated based on the measured LFs. 
We also elaborate on the robustness of our results.
In Sect. \ref{sec:discussion} we compare these to previous determinations of the UV LF and SFRD from the literature. 
In Sect. \ref{sec:conclusions} we finish with a summary and
present our conclusions.

We use the AB magnitudes system \citep{okegunn83} throughout and adopt
$\Lambda$CDM cosmology with $\rm{\Omega_m=0.3}$,
$\rm{\Omega_{\Lambda}=0.7}$ and $\rm{H_0=70\, km\, s^{-1}\,
  Mpc^{-1}}$.

\section{Data \& Samples}\label{sec:datasamples}
\subsection{The CFHT Legacy Survey Deep}
For this work we make use of publicly available data from the CFHTLS
Deep, a survey using MegaCam mounted at the prime focus of
  the CFHT which covers four independent fields of 1 square degree
each. Images are taken in the filters $ugriz$ and are pre-processed
using the Elixir system \citep{elixir04}. Image reduction is done
using the THELI pipeline \citep{erben05,2006A&A...452.1121H}, leading
to approximate $5\sigma$ point source limits of 27.5, 27.9, 27.9, 27.7
and 26.5 for the $ugriz$ bands, respectively. The limits for
  each of the fields lie within 0.2 mag from these average values.

Source Extractor \citep{bertinarnouts96} is used to create a
  multi-colour catalogue. Total $i$-band magnitudes are measured in
  Kron-like apertures \citep{kron80} using SExtractor's AUTO
  magnitudes. Every image is smoothed by convolution with a Gaussian
  filter to match the seeing of the image with the worst seeing value
  (typically the $u$-band). These corrections are typically small -
  all bands have a seeing below 1 arcsec. The dual-image mode of
  SExtractor is then used with the unconvolved $i$-band for source
  detection and isophotal magnitudes from the convolved bands to
  estimate colours.  An extensive description of the data reduction
  and catalogue creation is given in \citet{erben09} and
  \citet{hildebrandt09a}.

\subsection{$u$-, $g$-, and $r$-dropout samples}\label{sec:dropoutsamples}
Clean samples of $u$-, $g$-, and $r$-dropouts are selected based on the following selection
criteria \citep{hildebrandt09a}:
\begin{eqnarray}
u\rm{-dropouts} &:& 1.0 < (u-g) \land -1.0 < (g-r) < 1.2 \land \nonumber\\ &&1.5 \cdot(g-r) < (u-g) -0.75,\label{eqn:colourselectionu}\\
g\rm{-dropouts} &:& 1.0 < (g-r) \land -1.0 < (r-i) < 1.0 \land \nonumber\\ &&1.5 \cdot(r-i) < (g-r) -0.80,\label{eqn:colourselectiong}\\
r\rm{-dropouts} &:& 1.2 < (r-i) \land -1.0 < (i-z) < 0.7 \land \nonumber\\ &&1.5 \cdot(i-z) < (r-i) -1.00.\label{eqn:colourselectionr}
\end{eqnarray}

Furthermore, it is required that all LBGs have a $SExtractor$ CLASS\_STAR parameter of CLASS\_STAR $<$ 0.9, that $g$-dropouts are not detected in $u$, and that $r$-dropouts are neither detected in $u$ nor in $g$. Note that the colour selection criteria of the $u$-dropout sample in \citet{hildebrandt09a} are pretty conservative. We relaxed the $u-g$ cut slightly to make the selection more comparable to e.g. \citet{steidel03}, \citet{steidel2004} and \citet{st2}.

\begin{figure*}
\centering
\includegraphics[width=17cm]{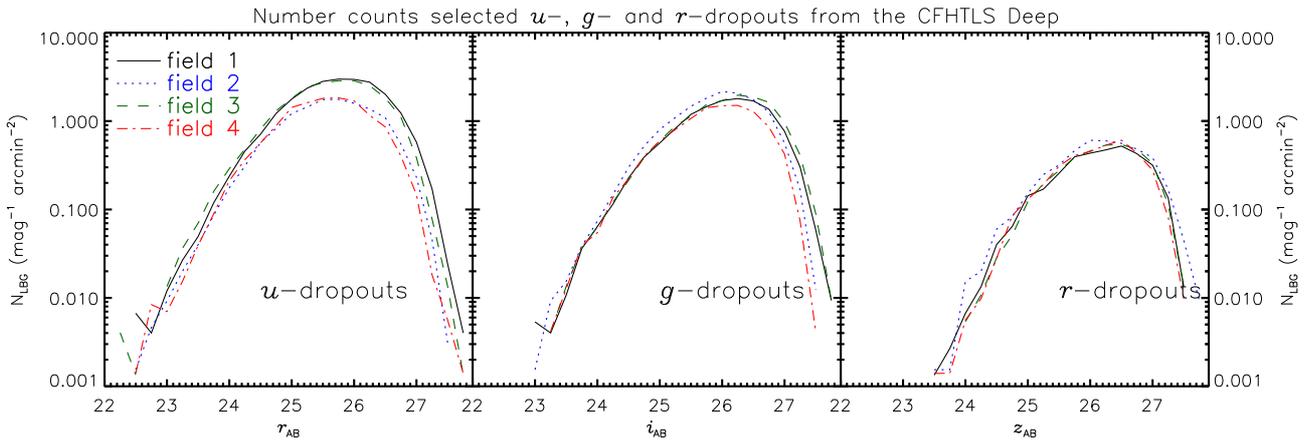}
\caption{From left to right the number counts of the \textit{u}-,
  \textit{g}-, and \textit{r}-dropouts in the CFHTLS Deep, as selected
  by the colour criteria of
  Eqs. \ref{eqn:colourselectionu}-\ref{eqn:colourselectionr}.}
\label{fig:nc_allplot}
\end{figure*}

This results in the selection of 50880 $u$-, 36226 $g$-, and 11411
$r$-dropouts in total over the four fields. Their magnitude
distributions are shown in Fig.~\ref{fig:nc_allplot}. Note the
differences in depth of the individual fields.

\subsection{Redshift distributions \& Contamination fractions}\label{sec:redshiftdist}
The majority of the selected sources is too faint to make a
spectroscopic redshift determination possible, and the brighter
candidates have not spectroscopically been observed yet. For this reason
\citet{hildebrandt09a} estimated the redshift distributions by means
of photometric redshifts and simulations. 

Throughout this paper we will use the mean redshift values from
simulations based on synthetic templates by \citet{bc93}, being $<z> \cong$ 3.1, 3.8 and 4.7 for the $u$-, $g$-, and $r$-dropouts respectively. We estimate
the uncertainty in the mean redshifts to 0.1 for the three dropout
samples.

\begin{figure*}
\includegraphics[width=17cm]{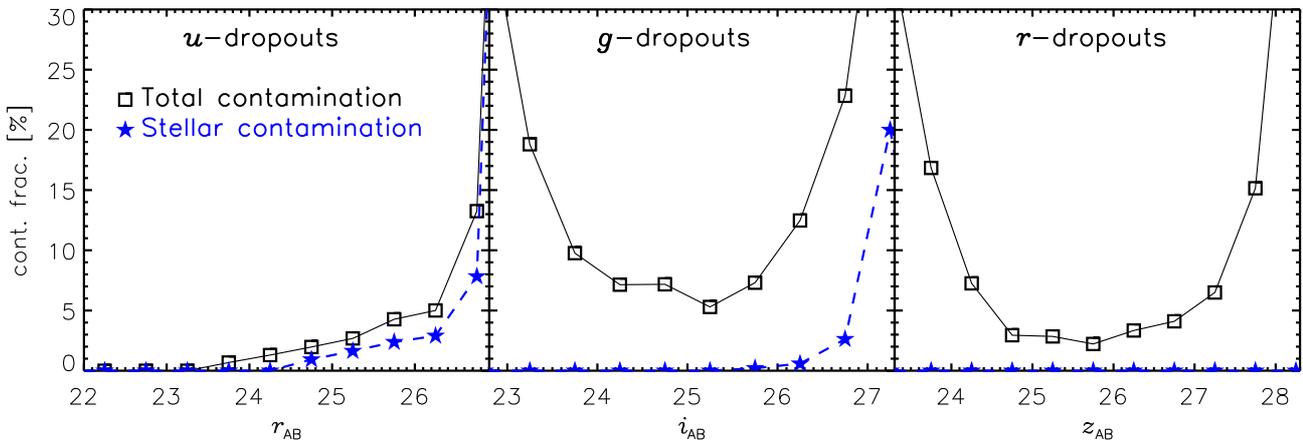}
\caption{Contamination fractions of stars and low-$z$ galaxies in the dropout samples. Blue $\star$-symbols connected by a dashed line show the stellar contamination fraction based on a galactic model. Black squares connected by the solid line show the total contamination fraction from \citet{hildebrandt09a}.}
\label{fig:contplot}
\end{figure*}

In order to address the amount of contamination in our LBG samples,
\citet{hildebrandt09a} consider the possibilities of stars and low-$z$
galaxies scattering in the selection boxes. Galaxies are simulated
based on templates from the library of \citet{bc93}, while the colours
of stars in the fields are estimated based on the TRILEGAL galactic
model \citep{girardi05}. Contamination fractions are shown graphically
in Fig.~\ref{fig:contplot}.

Stellar contamination is negligible for the brighter objects, as the
selection boxes steer away from the stellar locus. For faint objects
in the $u$-, and $g$-dropout samples, the stellar contamination
increases as a result of photometric scatter.

The contamination by low-$z$ galaxies is negligible in the $u$-dropout
sample, as the Lyman break for a $z\sim3$ galaxy is still blue-ward of
the $z\sim0$ Balmer/4000$\rm{\AA}$ break. For higher redshifts this
ceases to be the case, so that the $g$-, and $r$-dropout samples
suffer from a significant contamination fraction at the bright end,
where the LBG population is sparse. Faint low-$z$ objects are likely
to scatter into the selection box in each of the samples, so that the
contamination fractions are increased here.

Since the contamination fractions are low at the bright end of the $u$-dropouts, we have the potential to probe the LF to very low LBG galaxy densities. 
We inspect the 80 brightest objects ($r_{\rm{AB}}<23.2$) in the
$u$-dropout sample by eye and remove obvious spurious sources, 30 in total. A
strong, but certainly slightly subjective, rejection criterion is the
size of the region in which the flux is measured, i.e. the half-light
radius (13 objects rejected). Sources that are clearly blended by a bright neighbour
(2 objects), as well as sources that have a too large apparent size (3 objects). Also an asteroid track has been removed. 

In the $g$-, and $r$-dropout sample we do not probe
these low ($\lesssim 0.01\,\rm{mag^{-1}\,arcmin^{-2}}$) LBG galaxy densities,
since these points are unreliable due to the high contamination
fractions from low-$z$ objects (see Fig.~\ref{fig:contplot}). 
This prohibits a study of the bright end of the LF for the $g$-, 
and $r$-dropouts until the nature of the individual sources has been verified.

\section{Analysis - Survey completeness}\label{sec:analysis} \label{sec:completeness}
We use a detailed modelling approach to estimate the completeness of
the survey as a function of magnitude and redshift, for each of the
dropout samples. We add artificial objects to our images, with
colours representative of star-forming galaxies, and try to recover
them following the same source extraction and colour selection
criteria as for the real data. We investigate how the increasing
scatter in the colours for fainter objects influences the completeness
as a function of magnitude.  Furthermore, one expects that, for
fainter dropout objects, the redshift distribution of these objects
broadens due to the same effect. Hence we will also model this
as a function of redshift.

We describe our fiducial model SEDs, sizes and shapes of the simulated
objects below. The assumptions we make, are tested in
Sect. \ref{sec:robust} to estimate the robustness of the results.

\subsection{Model galaxies}\label{sec:modelgalaxies}
The \citet{bc93} stellar population synthesis library is used to set
up our fiducial galaxy SED model; a 100 Myr old galaxy template with constant star formation.
A \citet{millerscalo79} IMF is assumed. The optical depth
of neutral hydrogen, as a function of redshift, is modelled according to
\citet{madau95}. 

This template is reddened by the starburst extinction
law from \citet{calzetti00}, with a distribution in $E(B-V)$. This distribution we choose such that the UV-continuum slopes that we measure from the data are matched when we use our fiducial template as a base. To measure the UV-continuum slopes, we use a colour redward of 1600$\AA$ rest-frame, i.e. the $r-i$ colour for the $u$-dropouts and the $i-z$ colour for the $g$-dropouts. For the $r$-dropouts we can not perform a similar measurement because we do not have observations in a band redward of the $z$-band. Therefore we will use the same distribution of dust as we find for the $g$-dropouts.

\begin{figure}
\resizebox{\hsize}{!}{\includegraphics{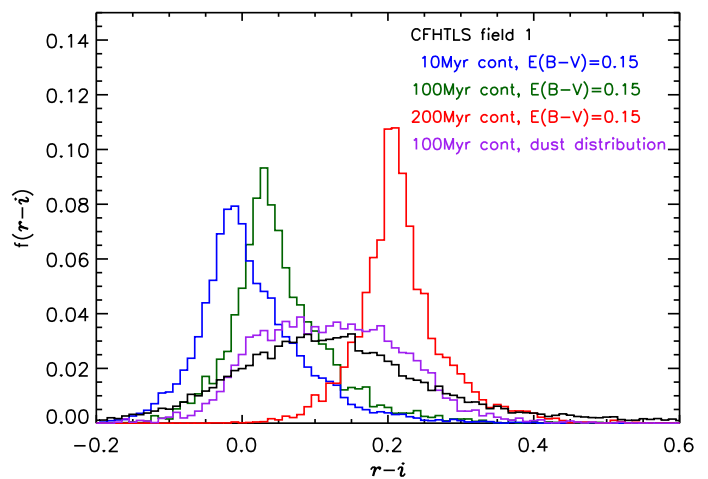}}
\caption{The distribution of UV-continuum slopes for the $u$-dropouts as measured by the $r-i$ colour from the CFHT data (black) compared with the outcomes from simulations. The colour is measured in a part of the spectrum without strong features, redward of 1600$\AA$. Both an older template and a higher amount of dust result in a redder UV colour. The amount of dust we need to add therefore depends on the age of the base template. For the $u$-dropouts we find that a uniform distribution of dust with 0.1$<E(B-V)<$0.4 gives a good fit to the data, when the template is 100 Myr old with a constant amount of star formation (purple).}
\label{fig:uvcontslopesu}
\end{figure}

For the $u$-dropouts we find that a uniform dust distribution with reddening between with 0.1$<E(B-V)<$0.4 gives a good fit to the data, see Fig. \ref{fig:uvcontslopesu}. For the $g$-dropouts we measure a larger spread in UV-continuum slopes, and find a reasonable fit when using a uniform dust distribution with reddening between 0.0$<E(B-V)<$0.5. It should be noted that the age and amount of dust attenuation of the template are highly degenerate, so that different combinations of these parameters fit the UV-continuum slopes in the data. It is especially important to correctly match the distribution of the UV-continuum slopes from the data with the model galaxies, since an increase (decrease) in the age of the galaxy model template has a similar effect on the LF as an increase (decrease) of the amount of dust. We will elaborate on this in Sect. \ref{sec:robust}. If we measure the UV-continuum slopes in our data for different magnitude bins, we do not find a significant evolution. Therefore we will use a distribution of dust attenuation in our simulations that is independent of magnitude.

LBGs have typical half-light radii of $r_{1/2} \sim$ 0.1"-0.3"
\citep{giavalisco96} and thus are unresolved by the CFHT and can be
treated as point sources. As the size of the PSF in the CFHT images is
strongly position dependent, we have to adapt the
injected sources accordingly. We parametrize the PSF by a Moffat
profile,
\begin{equation}
I =I_{0}\left[1+(2^{1/\beta}-1) \cdot \left(\frac{R}{R_{0}}\right)^2\right]^{-\beta}, 
\end{equation}
in which $\beta$ and $R_{0}$ are the parameters that we adapt to
adjust the shape and size of the profile respectively. $I_{0}$ represents the flux normalisation.

In order to measure the PSF as a function of position for the
different filters and fields, we first select several hundred stars
based on their magnitudes and half-light radii, in each field and each
filter. We measure the 50\% and 90\% flux radii of these stars,
$\rm{r_{50}}$ and $\rm{r_{90}}$, using SExtractor. The ratio of these
flux radii uniquely determines the Moffat-$\beta$ parameter, which, in
combination with either one of the flux radii, gives the Moffat
profile radius $R_{0}$ for each star. We find that the Moffat-$\beta$
parameter is fairly constant over the fields and filters $(\beta
\approx 4.0)$ and only $R_{0}$ changes significantly. To model $R_{0}$
as a function of position, we fit a 2-dimensional polynomial function
to the SExtracted values for $r_{50}$. This constrains the PSF size on
every position, for every field and filter.

We find that there is a $\sim$30\% difference in $R_{0}$ between the image center and boundaries. As this, in our ground-based wide-field survey, is by far the dominating effect in the apparent surface brightnesses of our sources, we do not assume an intrinsic size distribution for the sources in the main simulations of this paper.

\subsection{Eddington bias}\label{sec:eddingtonbias} 
If we consider a certain intrinsic magnitude
distribution of galaxies, the recovered distribution
after source extraction and colour selection will look different due
to statistical fluctuations in
the measurement. In our analysis we attempt to correct for this effect called
Eddington bias \citep{eddington13,  teerikorpi04}. It is hard to estimate this bias analytically
because the size of the magnitude scatter is an increasing function of
magnitude. Also, when approaching the completeness limit of the
survey, only the brighter objects will be detected. What generally
happens is that intrinsically faint objects will on average look
brighter than they are.

\begin{figure}
\resizebox{\hsize}{!}{\includegraphics{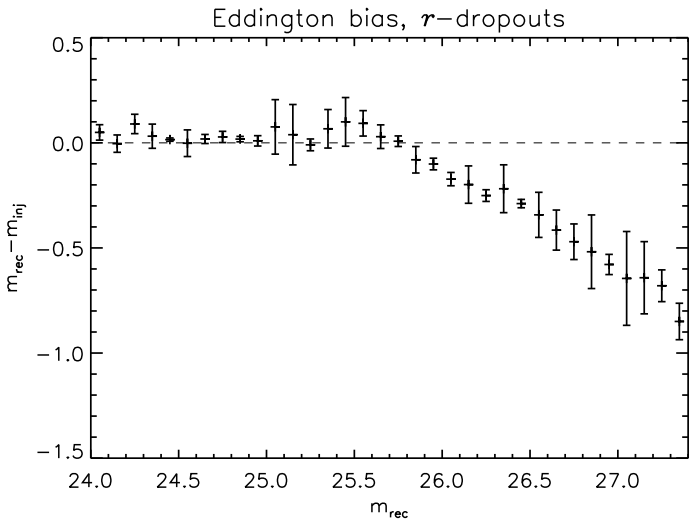}}
\caption{The difference between recovered and injected source
  magnitudes as a function of recovered magnitude for the
  $r$-dropouts. Similar trends appear in the other dropout
  samples. The error bars reflect the scatter from the four fields. A
  similar effect as shown around $\rm{m_{rec}}=25.3$, namely an increase
  of $\rm{m_{rec}-m_{inj}}$, is found by \citet[][Fig.19]{bouwens06},
  where observations were used rather than simulations.}
\label{fig:eddingtonbias}
\end{figure}

The Eddington bias can be estimated by comparing the intrinsic
magnitudes of the injected sources to the recovered magnitudes. Such a
comparison is shown in Fig.~\ref{fig:eddingtonbias}, stressing the
importance of that effect for faint magnitudes.

We want to inject sources with the same magnitude distribution as the
intrinsic distribution, to get an unbiased result. As the bias
is expected to be largest for fainter sources, it is especially
important to correctly model the faint-end of the intrinsic
distribution. Here we adopt a LF that is consistent with the deepest 
LBG survey, conducted by \cite{bouwens07}.

\subsection{Source injection and recovery} 
\begin{figure}
\resizebox{\hsize}{!}{\includegraphics{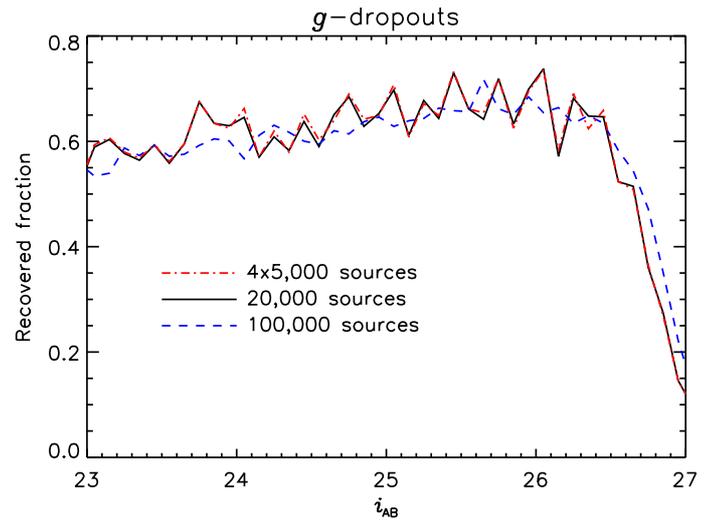}}
\caption{The recovered fraction of injected sources that also satisfy the $g$-dropout criteria, as a function of
  $i$-band magnitude, for 1 redshift step. Three curves are given for
  different amounts of injected sources, to see whether, and how, the
  presence of these sources influences the photometry of the
  analysis.}
\label{fig:consistencycheck2}
\end{figure}

\begin{figure}
\resizebox{\hsize}{!}{\includegraphics{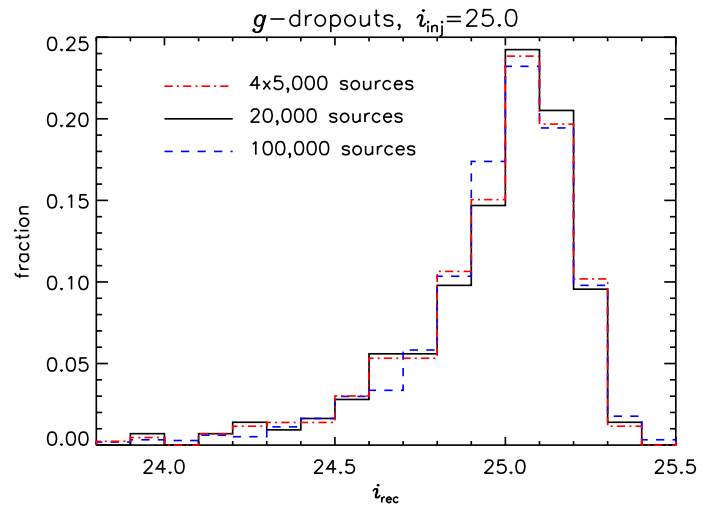}}
\caption{The distribution of measured source magnitudes for a
  population of injected simulated sources with intrinsic $i$-band
  magnitude=25.0, for 1 redshift step. Three curves are given for
  different amounts of injected sources, to see whether, and how, the
  presence of these sources influences the photometry of the
  analysis.}
\label{fig:consistencycheck3}
\end{figure}

Following 
this adopted intrinsic distribution we inject $20\,000$ sources in
each of the images, for 60 equal redshift steps between $z=0.0$ and
$z=6.0$. To verify that the injected sources do not significantly
influence each other by blending, nor that the background is
influenced significantly, we perform the following
tests. We inject the same $20\,000$ sources in 4 stages, 5000 sources
each, and do a third analysis where we inject $100\,000$ sources in
total. In Fig.~\ref{fig:consistencycheck2} the recovered fractions of
sources that also satisfy our $g$-dropout criteria are shown as a
function of magnitude, for one particular redshift step. Only for
faint magnitudes does the $100\,000$ curve deviate from the other ones,
which are identical in this regime. In
Fig.~\ref{fig:consistencycheck3} the distribution of recovered
{$g$-dropouts} with an intrinsic magnitude of $m=25.0$ is shown as a
function of recovered magnitude. We conclude that the injection of
$20\,000$ sources does not influence the images such that
the photometry would be perturbed significantly. A similar behaviour
is expected for the $u$- and $r$-dropout samples.


The clustering of LBGs is not taken into account. We assume this
effect to be insignificant for estimating completeness, as the
correlation length, which is typically around 5 Mpc
\citep{hildebrandt09a}, is very small compared to the survey
volume. Therefore we spread our simulated sources uniformly over the
images.

\subsection{Effective volumes}\label{sec:effvolumes}
Next we define the function $p(m,z)$ to be the number of sources
recovered with an observed magnitude in the interval $[m;m+\Delta m]$,
and are selected as dropouts, divided by the number of injected sources
with an intrinsic magnitude in the same interval
$[m;m+\Delta m]$ and a redshift in the interval $[z;z+\Delta z]$. 
Note that the definition of $p(m,z)$ is slightly different compared to the one used in e.g. \citet{st2}, as they do not take Eddington bias into account. In our definition, $p(m,z)$ could potentially be $>$ 1 as a result of this bias correction. 

The effective volumes ($V_{\rm{eff}}$) of our survey are given by
\begin{equation}
V_{\rm{eff}}(m)=A_{f}\int \frac{dV_{\rm{C}}}{dz}p(m,z)dz,
\end{equation}
where $A_{f}$ is the field area in square arcminutes, and
$\frac{dV_{\rm{C}}}{dz}$ is the comoving volume per square arcminute,
which depends on the adopted cosmology.

\begin{figure}
\includegraphics[width=6cm]{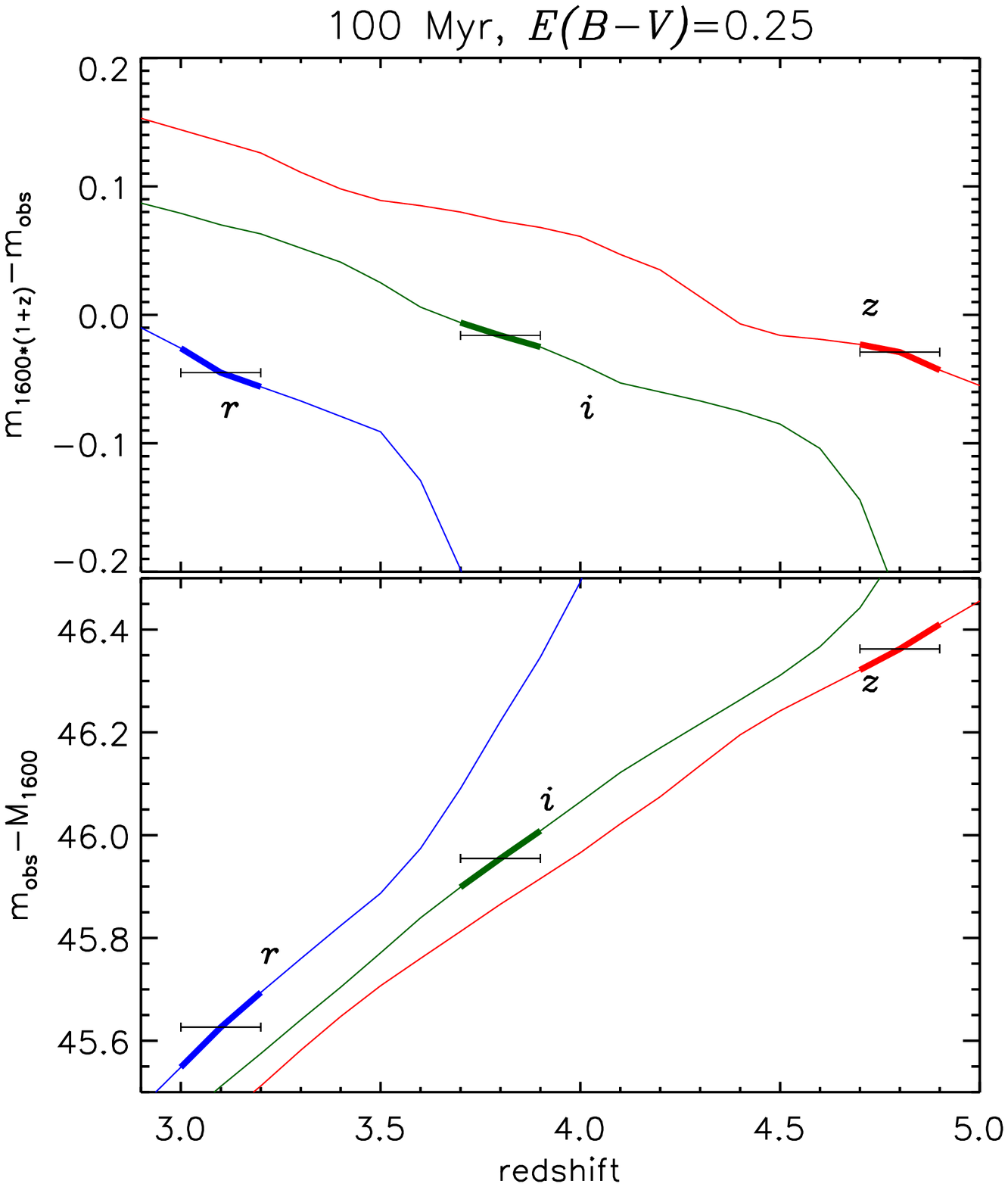}
\caption{Upper panel: k-correction to 1600$\rm{\AA}$ for the MegaCam
  $riz$ filters, as a function of redshift based on the 100 Myr old
  continuously star-forming template with a dust attenuation of $E(B-V)$=0.25. The
  average redshifts from the simulations and their uncertainties are
  represented by horizontal error bars. This leads to a corresponding
  error in the k-correction. Lower panel: Shifts from apparent
  magnitudes in the MegaCam $riz$ filters, to absolute magnitudes at
  1600$\rm{\AA}$. The Distance Modulus and k-correction are taken into
  account. The uncertainties in the average redshifts of the samples
  lead to uncertainties in the absolute magnitudes.}
\label{fig:kcorr}
\end{figure}

The magnitude is measured in the $r$-, $i$-, and $z$-bands for
  the $u$-, $g$-, and $r$-dropout samples, respectively. These bands
  probe flux at approximately 1600$\rm{\AA}$ rest-frame of the sources
  at the expected mean redshifts, so that only a minor k-correction
  will be sufficient to compare the results for the different epochs
  directly, see the upper panel of Fig.~\ref{fig:kcorr}. We transform
  the apparent magnitudes to absolute magnitudes and perform a
  k-correction to a rest-frame wavelength of 1600$\rm{\AA}$ using the mean redshifts for each of the dropout samples, i.e. we assume all sources to be at the same redshift.

The uncertainties in the mean redshifts\footnote{Note that we make use
  of two different redshift distributions in our analysis. To estimate
  both the k-correction and the effective volumes we use the
  distribution given by the simulations described in
  Sect. \ref{sec:completeness}, while we use the distribution from the
  simulations from \citet{hildebrandt09a} to shift the LF in the
  magnitude direction. The redshift distribution from the latter
  simulations are expected to be more reliable because they take a
  wide variety of template galaxy models into account, but we cannot
  use them for effective volumes \& k-corrections because of
  computational constraints. Rather we have to simulate those with a
  single template.}  are expected to be approximately $0.1$ for the
three dropout samples, as argued in Sect. \ref{sec:redshiftdist}. This
leads to uncertainties in both distance modulus and k-correction,
resulting in a systematic error in the absolute magnitudes of our
estimated LF, see the lower panel of Fig.~\ref{fig:kcorr}. The final
uncertainties are about 0.07, 0.05 and 0.04 in the absolute magnitude
for the $u$-, $g$-, and $r$-dropout samples, respectively.

\section{Results}\label{sec:results}
\subsection{The UV Luminosity Function at $z$=3-5}
After dividing the number counts by $V_{\rm{eff}}$, which corrects for
incompleteness and Eddington bias, and subtracting the distance modulus and the
k-correction from the apparent magnitudes, we obtain the LF in
absolute magnitudes at 1600$\AA$.

\begin{figure*}
\includegraphics[width=17cm]{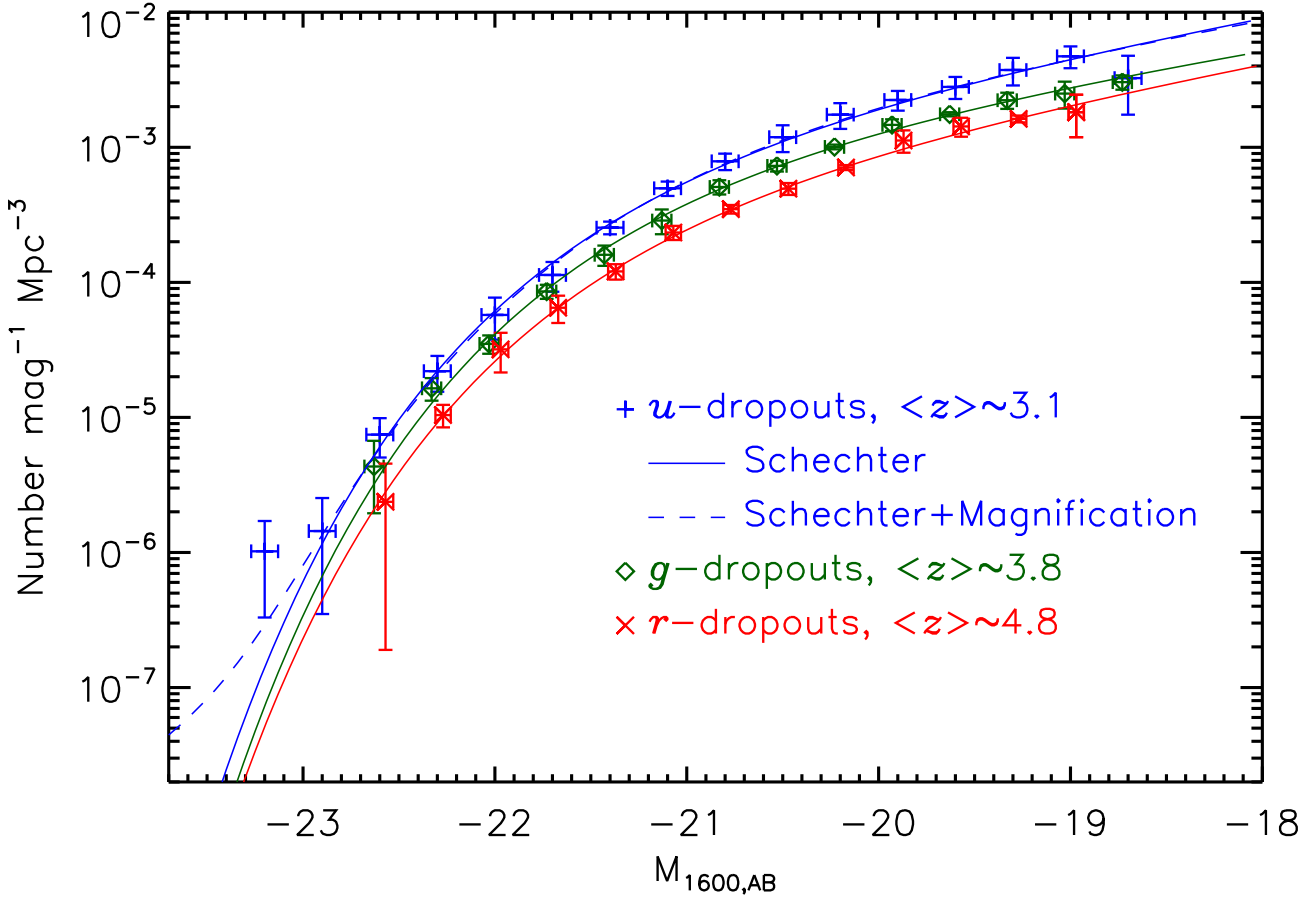}
\caption{The LFs of LBGs in the CFHTLS-Deep fields. Data points and
  best-fitting Schechter functions are shown for the $u$(blue)-,
  $g$(green)-, and $r$(red)-dropouts. For legibility we applied a small offset on the x-axis values of the $g$-, and $r$- dropouts. The dashed blue curve shows the best fitting Schechter function for the $u$-dropouts after magnification effects have been included, as described in Sect.~\ref{sec:magnification}.}
\label{fig:newLFme}
\end{figure*}

Results from the four fields are binned to $\Delta$mag=0.3, combined,
and shown in Fig.~\ref{fig:newLFme} and Table
\ref{tab:datame}. Uncertainties in the magnitude direction are due to
uncertainties in the redshift distribution of source galaxies.  The
four independent analyses of the fields allow us to estimate
field-to-field variations for each of the data points. Vertical error
bars reflect either this uncertainty, or the Poisson noise term,
whichever is largest. Usually the field-to-field variance
dominates. As a consequence of the way these are computed, Poisson
noise is always taken into account.

\begin{table}
\caption{The estimated LFs from the CFHTLS Deep.}
\label{tab:datame}
\centering
\begin{tabular}{c|ccc}
\hline\hline
 & $u$-dropouts & $g$-dropouts & $r$-dropouts\\
 &$\phi_{k}\,[10^{-3}$&$\phi_{k}\,[10^{-3}$&$\phi_{k}\,[10^{-3}$\\
$M_{\rm{1600,AB}}$ & $\rm{Mpc^{-3}\,mag^{-1}}]$ &$\rm{Mpc^{-3}\,mag^{-1}}]$ &$\rm{Mpc^{-3}\,mag^{-1}}]$\\
\hline
-23.20 & 0.001 $\pm $ 0.001 &         -          &          -         \\
-22.90 & 0.001 $\pm $ 0.001 &         -          &          -         \\
-22.60 & 0.007 $\pm $ 0.002 & 0.004 $\pm $ 0.002 & 0.002 $\pm $ 0.002 \\ 
-22.30 & 0.022 $\pm $ 0.007 & 0.016 $\pm $ 0.003 & 0.010 $\pm $ 0.002 \\ 
-22.00 & 0.057 $\pm $ 0.020 & 0.035 $\pm $ 0.005 & 0.032 $\pm $ 0.010 \\ 
-21.70 & 0.113 $\pm $ 0.028 & 0.086 $\pm $ 0.010 & 0.065 $\pm $ 0.015 \\ 
-21.40 & 0.254 $\pm $ 0.027 & 0.160 $\pm $ 0.027 & 0.121 $\pm $ 0.016 \\ 
-21.10 & 0.497 $\pm $ 0.061 & 0.287 $\pm $ 0.060 & 0.234 $\pm $ 0.028 \\ 
-20.80 & 0.788 $\pm $ 0.110 & 0.509 $\pm $ 0.061 & 0.348 $\pm $ 0.025 \\ 
-20.50 & 1.188 $\pm $ 0.267 & 0.728 $\pm $ 0.067 & 0.494 $\pm $ 0.050 \\ 
-20.20 & 1.745 $\pm $ 0.377 & 1.006 $\pm $ 0.040 & 0.708 $\pm $ 0.030 \\ 
-19.90 & 2.240 $\pm $ 0.373 & 1.465 $\pm $ 0.147 & 1.123 $\pm $ 0.211 \\ 
-19.60 & 2.799 $\pm $ 0.519 & 1.756 $\pm $ 0.063 & 1.426 $\pm $ 0.229 \\ 
-19.30 & 3.734 $\pm $ 0.863 & 2.230 $\pm $ 0.305 & 1.624 $\pm $ 0.095 \\ 
-19.00 & 4.720 $\pm $ 0.866 & 2.499 $\pm $ 0.564 & 1.819 $\pm $ 0.630 \\ 
-18.70 & 3.252 $\pm $ 1.508 & 3.038 $\pm $ 0.370 &          -         \\ 
\hline
\end{tabular}
\end{table}

We fit a Schechter function \citep{schechter76} to the binned data points, 
\begin{equation}\label{eqn:schechterfunc}
\phi(M)dM=0.4 \ln(10) \phi^{*} 10^{0.4(\alpha+1)(M^{*}-M)} \exp(-10^{0.4(M^{*}-M)}),
\end{equation}
with $M^{*}$ being the characteristic magnitude, $\alpha$ being the
faint-end slope, and $\phi^{*}$ being the overall normalisation.

\begin{figure}
\resizebox{\hsize}{!}{\includegraphics{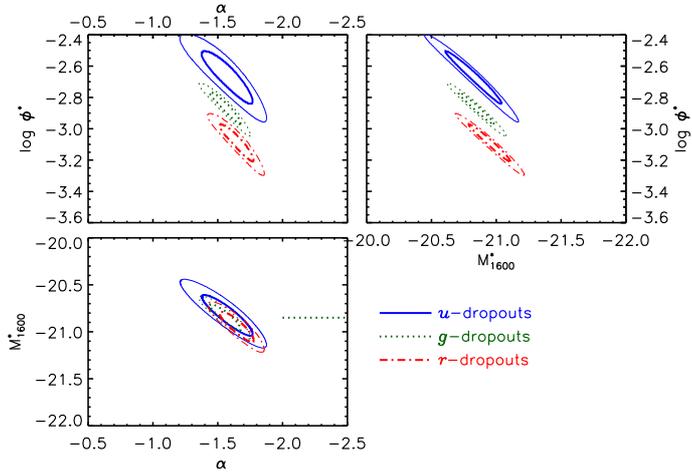}}
\caption{The 68\% and 95\% likelihood contours for different Schechter
  parameter combinations. Shown are the results for the $u$-dropouts
  at $z\sim3$ (blue, solid), the $g$-dropouts at $z\sim4$ (green,
  dots), and the $r$-dropouts at $z\sim5$ (red, dash-dots). The $u$-dropout contours represent the best Schechter parameters when we include magnification effects, as described in Sect.~\ref{sec:magnification}.}
\label{fig:ellipsallme}
\end{figure}

Using $\chi^{2}$ statistics on a three dimensional grid of
  $500^{3}$ different Schechter parameter combinations, we find the
  minimal value ($\chi_{\rm{min}}^{2}$) for each of the dropout
  samples yielding the best fit values.  To estimate the errors in the
  fitted parameters, we project the 3-dimensional distribution of
  $\chi^2$ to 3 planes by taking the minimum $\chi^{2}$ along the
  projected dimension. In Fig. \ref{fig:ellipsallme} the 68.3\% and
  95.4\% confidence levels are shown, which correspond to a $\Delta
  \chi^{2}$=2.3 and 6.17 with respect to $\chi^2_{\rm min}$. In Table
  \ref{tab:sparms} we give the 68.3\% confidence levels on
  each individual parameter, corresponding to $\Delta
  \chi^{2}$=1.0. 

Note, however, that this error estimate assumes Gaussian errors, and that the errors on
the data points are independent. Especially for the $u$-dropouts this
is probably not the case. The normalisation of the LF seems to be
systematically slightly different for each of the fields (see also Fig. \ref{fig:nc_allplot}), giving a systematic
uncertainty in $\phi^{*}$ of about 30\%. For the $u$-dropouts this
effect is largest because of a slightly more uncertain flux
calibration in the $u$-band compared to the $g$- and $r$-bands. The
effective filter throughput changes with time in the UV as the
atmosphere is changing, and also the camera is less sensitive in this
wavelength regime resulting in larger shot-noise.

\begin{table}
\caption{A comparison between the best fitting Schechter parameters
  and their 68\% confidence intervals for the $u$-, $g$-, and
  $r$-dropouts.}
\label{tab:sparms}
\begin{center}
\begin{tabular}{ccccc}
\hline\hline
&& $\phi^{*}$ &&\\
Sample &$\rm{M}_{\rm{UV}}^{*}$ $^{\mathrm{a}}$&[$10^{-3}\, \rm{Mpc}^{-3}$]&$\alpha$&$\chi^2/\rm{dof}$\\
\hline
$u$&$-20.94^{+0.14}_{-0.13}$&$1.79^{+0.51}_{-0.38}$&$-1.65^{+0.12}_{-0.11}$&$0.52$\\
$g$&$-20.84^{+0.09}_{-0.09}$&$1.36^{+0.23}_{-0.20}$&$-1.56^{+0.08}_{-0.08}$&$0.36$\\
$r$&$-20.94^{+0.10}_{-0.11}$&$0.83^{+0.15}_{-0.14}$&$-1.65^{+0.09}_{-0.08}$&$0.19$\\
$u^{\mathrm{b}}$&$-20.84^{+0.15}_{-0.13}$&$2.11^{+0.63}_{-0.45}$&$-1.60^{+0.14}_{-0.11}$&$0.41$\\
\hline
\end{tabular}
\end{center}
\begin{list}{}{}
\item[$^{\mathrm{a}}$] Due to uncertainties in the redshift
  distributions there is an additional error component of 0.07, 0.05,
  and 0.04 in the estimated $\rm{M}_{\rm{UV}}^{*}$ for the $u$-, $g$-,
  and $r$-dropout samples, respectively.
 \item[$^{\mathrm{b}}$] Best-fitted Schechter parameters for a model where the function is modified with a magnification distribution.  
\end{list}
\end{table}

\subsubsection{Magnification contribution at low densities}\label{sec:magnification}
Due to inhomogeneities in the matter distribution between distant sources and the observer, paths of photons get slightly perturbed. This results in a distortion of the shape and a magnification of distant sources. When a source is magnified by a factor $\mu$, the flux gets boosted by the same amount. One can relate an intrinsic luminosity distribution to an observed distribution if the magnification distribution is known, as was shown by \citet{jainlima2010}. \citet{hilbert2007} estimate magnification distributions for different source redshifts by shooting random rays through a series of lens planes created from the Millennium Simulation. The width of the magnification distribution is found to increase with increasing source redshift, and the peak position of the distribution decreases slightly with increasing source redshift.

Magnification can account for a strong deviation from a Schechter function where the slope of the intrinsic luminosity distribution is very steep, see \citet{jainlima2010}. We stress again that we measure the LF from a volume that is much larger than has been used before. The bright end of our \mbox{$g$-,} and $r$-dropout samples suffers from increasing amounts of contamination. Only for the $u$-dropouts we probe the distribution of $u$-dropouts at the bright end down to a density of $10^{-6}\,\rm{mag}^{-1}\,\rm{Mpc}^{-3}$. We use the magnification distribution for a source redshift of $z=3.1$ that was kindly provided by Stefan Hilbert to improve our model to the data. 

Writing the LF as a function of magnitude, we use the following equation to correct the Schechter function for magnification effects. It is equivalent to the expression used by \citet{jainlima2010}.
\begin{equation}\label{eqn:magnification}
\phi(m_{obs})=\int d\mu \, P(\mu) \, \phi^{*}(m_{obs}+2.5 \log(\mu)), 
\end{equation}
where $\phi^{*}$ is the Schechter function defined by Eq. \ref{eqn:schechterfunc}. The new function yields a slightly better fit to the bright end of the LF, reducing the formal $\chi^2/\rm{dof}$ from 0.52 to 0.41. Find the new Schechter parameters, together with their 68.3\% confidence levels in Table \ref{tab:sparms}, and their 2-dimensional 68.3\% and 95.4\% confidence contours plotted in Fig.~\ref{fig:ellipsallme}. The best fitting function is the dashed curve in Fig.~\ref{fig:newLFme}. 

As the bright selected $u$-dropouts are likely to be significantly magnified, we expect them to appear close to a massive foreground galaxy or group of galaxies that acts as a lens. We inspected the brightest ($r_{AB}<$23.2) $u$-dropouts by eye and find that this is indeed the case for many of them. Note that the model is still uncertain as the Millennium simulation does not include baryonic matter, and assumes $\sigma_{8}=0.9$ \citep{springel2005}, where recent estimates indicate a lower value around 0.8 \citep{komatsu2010}. Also, in the magnification probability distribution the possibility of multiply imaged systems is ignored. To rule out contamination in the LBG sample at the bright end of the LF, the nature of each bright object has to be verified spectroscopically.

A statistically much better sample can be selected from the CFHT Legacy Survey Wide, consisting of 170 square degree imaging in $ugriz$ of shallower depth. We leave this for future studies.

\subsection{The UV Luminosity Density and SFR Density}
Next we estimate the UV luminosity density (UVLD) at the different epochs. To
be able to compare our results with the results from previous studies,
we will integrate the data points down to $L=0.3L^{*}_{z=3}$, where
$L^{*}_{z=3}$ $(\rm{=9.4\times10^{28}\, erg\, s^{-1}\, Hz^{-1}\, at\, 1600\AA})$ is the characteristic luminosity of our $u$-dropout
sample. Results are shown in the odd-numbered rows of Col.~3 in Table~\ref{tab:sfrd}. However, for steep faint-end slopes of $\alpha<-1.6$, more
than 50\% of the UV luminosity is expected to be emitted by lower
luminosity sources. Therefore we will make a second estimate of the UV
luminosity density by integrating the best-fitting Schechter function
over all luminosities. This results in
\begin{equation}
\rho_{\rm{UV}}=\phi^{*}L^{*}\Gamma(\alpha+2),
\end{equation}
where $\Gamma$ is Euler's Gamma function.  Although this full
integral of the LF depends strongly on uncertainties in the faint-end
slope, we use it to provide an upper limit to the UVLD. The results
are shown in the even-numbered rows of Col.~3 in Table~\ref{tab:sfrd}.

The effective extinction in the UV is a strong function of the amount of dust. At these high redshifts ($z \gtrsim 3$) the only estimate for the amount of dust is based on a measure of the UV continuum slope. Note however that there is a strong degeneracy between the age and the amount of dust in the template if the rest-frame IR is not covered, see e.g. \citet[][]{papovich2001}.

\citet{bouwens09} recently measured the UV-continuum slope of LBGs at
high redshifts from deep HST data, from which the amount of dust obscuration could be
estimated as a function of LBG magnitude. The values they find at the characteristic magnitudes of our samples are $E(B-V)=0.15$ for $z=3,4$,
and $E(B-V)=0.10$ for the $z\sim5$ sample. \citet{bouwens09} find a decreasing amount of dust for fainter magnitudes. For consistency we will use the relationships they estimate to correct for dust extinction in our data, and not the values from Sect.~\ref{sec:modelgalaxies}.

\citet{meurer1999} find a relation between the UV-continuum slope and the extinction by dust. \citet{bouwens09} use this relation and find, upon integrating down to $L=0.3L^{*}_{z=3}$, density correction factors of $6.0^{+1.8+2.1}_{-1.4-1.6}$, $5.8^{+0.8+2.1}_{-0.7-1.5}$, and $2.7^{+0.7+1.0}_{-0.5-0.7}$ for the three redshift samples, respectively. Both random errors and systematic errors are quoted \citep{bouwens09}.

We now convert the UV luminosity density into the star
formation rate density, $\rho_{\rm{SFR}}$, at the different epochs
using \citep{madau98},
\begin{equation}
  L_{UV}=8.0 \times 10^{27} \, \left(\frac{\rm{SFR}}{M_{\sun}\rm{yr^{-1}}}\right) \, \rm{erg\,s^{-1}\,Hz^{-1}}.
\end{equation}
This relation assumes a 0.1-125$\rm{M}_{\sun}$ Salpeter IMF and a constant star formation rate of $\gtrsim$ 100 Myr. The
resulting estimates of $\rho_{\rm{SFR}}$ are shown in Col. 4 of Table
\ref{tab:sfrd}, where we have corrected for dust
extinction. In Sect. \ref{sec:SFRDcomparison} we compare these
  estimates to values reported in previous studies.

\begin{table}
\caption{ The $\rho_{\rm{UV}}$ and $\rho_{\rm{SFR}}$ for the different
  dropout samples. The first lines for each sample correspond to sums
  over the data points down to $L = 0.3L^{*}_{z=3}$ while the second
  lines correspond to integrals over the best-fit Schechter
  functions.}
\label{tab:sfrd}
\begin{center}
\begin{tabular}{cccc}
\hline\hline
Sample &Integral limit & $\rho_{\rm{UV}}$ $[10^{26}$ & $\rho_{\rm{SFR}}$\\
&&$\rm{erg\,s^{-1}\,Hz^{-1}\,Mpc^{-3}}]^{\mathrm{a}}$&$[\rm{M_{\sun}\,yr^{-1}\,Mpc^{-3}}]^{\mathrm{a,b}}$\\
\hline
$u$&$L > 0.3L^{*}_{z=3}$&$1.73\pm0.09$&$0.129^{+0.064}_{-0.036}$\\
&Schechter&$4.41$&0.154\\
$g$&$L > 0.3L^{*}_{z=3}$&$1.07\pm0.03$&$0.078^{+0.032}_{-0.019}$\\
&Schechter&$2.62$&0.092\\
$r$&$L > 0.3L^{*}_{z=3}$&$0.80\pm0.03$&$0.027^{+0.013}_{-0.007}$\\
&Schechter&$2.19$&0.038\\
\hline
\end{tabular}
\end{center}
\begin{list}{}{}
\item[$^{\mathrm{a}}$] Due to uncertainties in the redshift
  distributions, there is an additional error component of $\sim$7\%,
  $\sim$5\%, and $\sim$4\% in the estimated $\rho_{\rm{UV}}$ and
  $\rho_{\rm{SFR}}$ values for the $u$-, $g$-, and $r$-dropout
  samples, respectively.
\item[$^{\mathrm{b}}$] Corrected for dust extinction using the luminosity dependent correction factors from \citet{bouwens09}. Systematic errors as a result from the age-dust degeneracy are also included.
\end{list}
\end{table}

Note, however, that some sources like AGN, which might be included in our dropout samples, add to the total UV
luminosity density in the Universe, though do not contribute to the
SFRD.

\subsection{Robustness of our results} \label{sec:robust}
Our fiducial template model is a 100 Myr old continuously star-forming galaxy with a uniform distribution of dust centred around $E(B-V)$=0.25. This dust distribution was chosen such that the distribution of UV-continuum slopes of the recovered simulated sources matches the distribution of UV-continuum slopes in the real data (see Fig. \ref{fig:uvcontslopesu}).
We test some of the assumptions we made in
  Sect.~\ref{sec:modelgalaxies} by checking their influence on the
  final LFs. 

\begin{figure*}
\centering
\includegraphics[width=16.5cm]{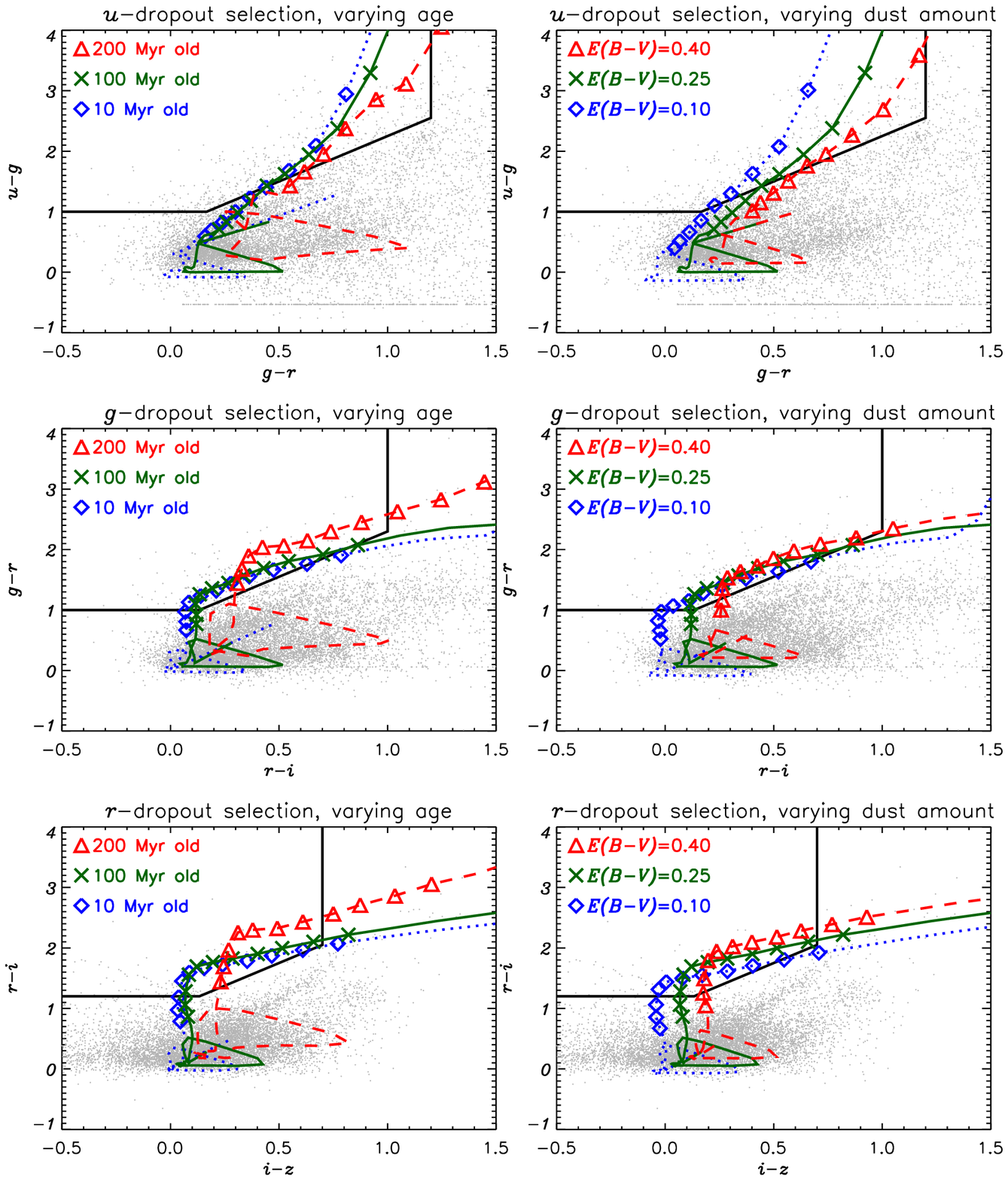}
\caption{The grey points represent the colours of 10,000 objects in
  field 1 of the CFHTLS Deep. The black boxes are defined by
  Eqs. \ref{eqn:colourselectionu}-\ref{eqn:colourselectionr} and are
  used to select $u$-(top panels), $g$-(middle panels), and
  $r$-(bottom panels) dropouts. The coloured tracks represent the
  colours of a template galaxy as a function of $z$. They are
  evaluated at intervals of $\Delta z$=0.1. The symbols mark redshifts
  from 2.5 to 3.5, from 3.2 to 4.2 and from 4.2 to 5.2 for $u$-, $g$-,
  and $r$-dropouts respectively. The green curve with $\times$-symbols represents a reference model, a 100 Myr old continuously star-forming template with a dust attenuation of
  $E(B-V)$=0.25. In the left panels we consider redder (bluer)
  templates by increasing (decreasing) the template age, see the red
  (blue) curves and the $\triangle$ ($\Diamond$)-symbols. In the right panels we consider redder (bluer)
  templates by increasing (decreasing) the amount of dust in the
  template, see the red (blue) curves and the $\triangle$ ($\Diamond$)-symbols. Note that we use a distribution of dust in our fiducial analysis (see Sect. \ref{sec:modelgalaxies}).}
\label{fig:colortrackall}
\end{figure*}

As a reference SED we use a 100 Myr old galaxy model with constant star formation and a single dust attenuation value of $E(B-V)$=0.25. We consider redder (bluer) templates by either increasing (decreasing)
the age of the star-forming period, or increasing (decreasing) the amount of
dust. In Fig.~\ref{fig:colortrackall} the colours of these alternative
templates, as they would be measured by the MegaCam $ugriz$ filter
set, are shown as a function of redshift. As the quality of the
Schechter fit is high in all cases ($\chi^{2}/\rm{dof} < 1.0$), we
present the differences by comparing the Schechter parameters:

\begin{itemize}
\item The faint-end slope $\alpha$ depends on the colour of the
  spectral template. In the $g$-, and $r$-dropout samples, redder templates tend to give steeper $\alpha$
  than bluer templates. The reason for this is as follows, with the
  $g$-dropouts as an example. Sources in the selection box have red observed $g-i$ colours. For faint magnitudes (note that the reference magnitude is measured in the
  $i$-band), the $g$-band magnitude exceeds the limiting magnitude of the survey. Only a lower limit on $g-r$ can then be given and the source moves out of the dropout selection box. The magnitude at which this happens depends on the $g-i$ colour of the template. As the average $g-i$ colour in the selection box is redder for red templates (Fig.~\ref{fig:colortrackall}), the detection rate of red sources is suppressed at faint magnitudes. This argument holds for any red template in the $g$-, and $r$-dropout samples. Fig.~\ref{fig:colortrackall} indicates that the opposite effect happens in the $u$-dropout sample as the $u-r$ colour is generally bluer (redder) for redder (bluer) templates in the selection box. This effect is indeed also inferred from the simulations.
There is an additional effect due to the requirement that $g$-dropouts are not detected in $u$, and that $r$-dropouts are neither detected in $g$ nor in $u$. This suppresses the detection of bright $g$-, and $r$-dropouts, especially at low redshifts. The bluer the $u-i$ colour for the $g$-dropouts (i.e. the bluer the template), the stronger this effect is. A similar argument holds for the $r$-dropouts. Therefore the $V_{\rm{eff}}$'s are higher in the faint
  magnitude regime, so that the LFs are lower at this end. The
  ranges of best-fitted $\alpha$ values, for the template spectra we
  considered, are $-1.82 < \alpha <-1.38$ for the $u$-, $-1.82 <
  \alpha <-1.42$ for the $g$-, and $-2.14 < \alpha <-1.40$ for the
  $r$-dropout sample.
  
  \item The characteristic magnitude, $M^{*}$, does not sensitively depend
  on the template spectrum chosen. The ranges of best-fitted $M^{*}$ values, for the template spectra we
  considered, are $-20.88 < M^{*} <-20.72$ for the $u$-, $-21.06 <
  M^{*} <-20.74$ for the $g$-, and $-21.26 < M^{*} <-20.88$ for the
  $r$-dropout sample. However, another systematic
  uncertainty in this parameter is due to the unknown redshift
  distribution of source galaxies, see Fig. \ref{fig:kcorr}, which
  depends on the mix of templates used by \citet{hildebrandt09a}.
    
\item The normalisation $\phi^{*}$ decreases (increases) when the faint-end slope becomes steeper (shallower). The best-fit Schechter parameters move then in the direction of the degeneracy of the ellipse in the upper left part of Fig.~\ref{fig:ellipsallme}. The
  ranges of best-fitted $\phi^{*}$ values  [$10^{-3} \,\rm{Mpc}^{-3}$], for the template spectra we
  considered, are $1.36 < \phi^{*} <2.89$ for the $u$-, $0.76 <
  \phi^{*} <1.74$ for the $g$-, and $0.39 < \phi^{*} <1.12$ for the
  $r$-dropout sample.
\end{itemize}

Some studies \citep[e.g.][]{st2} make use of a starburst template instead of a continuously star-forming model. The stellar population in a starburst template is older on average, and therefore the colours will be redder. However, for a template age of 100 Myr the difference in colours is very small. We compare the Schechter parameters that we measure after using our reference model (i.e. a 100 Myr continuously star-forming template with a dust reddening of $E(B-V)$=0.25) with a model where we change the star-formation law to a starburst. We find the Schechter parameters to change in the directions that are expected for a redder template, as explained above. However the differences are insignificant since they are much smaller than the statistical errors on the Schechter parameters.

We stress again that we use a mix of dust amounts in our standard analysis to match the UV-continuum slope distribution that is measured from the data. Especially for the $u$-, and $g$-dropouts this puts strong constraints on the combination of the age and the amount of dust in the model template, so that we can reduce the systematic error to a minimum.

To justify the assumptions we make regarding the shapes of our
simulated sources we also estimated the systematic error on the LF due
to this component. Because we expect similar results in the three
dropout samples, we only run these simulations for the
$g$-dropouts. We inject sources that are 0.05'' larger 
than the measured position-dependent PSF, and
compare values of Moffat parameter $\beta=3.0$ and $\beta=5.0$ with
our fiducial $\beta=4.0$ parameter. We find the following:

\begin{itemize}
\item As expected, $\alpha$ becomes steeper for more extended sources
  (i.e. increasing the $R_{0}$ or decreasing $\beta$), as this causes
  the peak surface brightness to drop. This only significantly affects the $V_{\rm{eff}}$'s at the faint end of the LF. Estimations of $\alpha$ range
  from $-1.94 < \alpha <-1.36$. A similar change is expected for the
  $u$-, and $r$- dropout studies.
The other Schechter parameters, $\phi^{*}$ and $M^{*}$, then also change slightly as they move in the direction of the degeneracy in Fig.~\ref{fig:ellipsallme}. 

\end{itemize}

Furthermore we find that, based on the different template spectra, the
estimated UV luminosity density varies. The ranges are, upon
integration down to $L<0.3L^{*}_{z=3}$, in units $\rm{[10^{26} erg\,s^{-1}\,Hz^{-1}\,Mpc^{-3}}]$, 
$1.30< \rho_{\rm{UV}} <1.98$ for the $u$-, $0.92< \rho_{\rm{UV}} <1.15$ for the $g$-, and $0.80< \rho_{\rm{UV}} <0.92$ for the $r$-dropouts.

In order to illustrate the effect that the Eddington bias can have on
LF estimations we repeat our analysis with slightly changed
$p(m,z)$. In the completeness simulations we bin the recovered sources
by their intrinsic magnitude instead of their recovered
magnitude. Doing so leaves the Eddington bias uncorrected. We find that the Schechter fit to the resulting LFs are not as good as for our standard analysis, especially for the $g$-, and $r$-dropouts, where we find that $\chi_{\rm{min}}^{2}/\rm{dof}$ would be 1.0 and 3.0, respectively. For the $u$-dropouts the best-fitting parameters are not changed significantly, but for the $g$-, and $r$-dropouts the faint-end slope steepens while the normalisation of the Schechter function decreases.

Note that we did not account for the presence of Lyman- $ \alpha $ emission in our simulations, although this line clearly contributes to broadband fluxes. 
\citet{shapley2003} measured the contribution of this line in the spectra of $ z \sim 3 $ LBGs and concluded that only $ \sim 25 \% $ of the sample showed significant Ly $ \alpha $ emission such that EW(Ly$\alpha$) $> 20 \AA$, see also \citet{reddy2008}. Although the relative contribution of this line is thought to increase towards higher redshift and fainter continuum luminosity, this has not been quantified yet. We will therefore describe possible biasses due to the presence of this line in the measurement of the LF only qualitatively.

If the line appears in emission, it contributes to the flux in the CFHT $g$-band for redshifts of about $2.5<z<3.5$, in the $r$-band for redshifts of about $3.7<z<4.6$, in the $i$-band for redshifts of about $4.8<z<5.9$, and in the $z$-band for redshifts of about $z>5.8$. Following the redshift distributions from \citet{hildebrandt09a}, we expect that the line predominantly contributes to the middle band in a two colour selection for the $u$-, and $r$- dropout samples, moving the source to the upper left in Fig. \ref{fig:colortrackall}. This effect causes the effective volumes to rise, thereby lowering the LF measurement. If the line indeed gets stronger at faint magnitudes, this would bias the LF measurement and results in a shallower alpha. In our $g$-dropout sample the line is expected to contribute to the flux in both the $r$- and $i$-band, depending on the redshift of the particular source. In the case where the line falls in the $r$-band, the effective volumes rise and the LF we measured is too high. For the redshifts where the line falls in the $i$-band, the sources would move to the right in Fig. \ref{fig:colortrackall} so that the effective volumes are decreased and the LF increases. To estimate which effect prevails, the redshift distribution needs to be measured spectroscopically. Some objects show Ly-$\alpha$ in absorption rather than emission, which would give the opposite effect.

Note that \citet{bouwens07} model the contribution of Ly$\alpha$ in the measurement of their LF, by using a simple model where 33\% of the $z \sim 4-5$ LBGs have EW(Ly $\alpha$ )= 50 $\AA$, independent of the continuum luminosity. They find that this affects the normalisation of the LF by only $\sim 10\%$. However, we stress again that the measure of $\alpha$ might be biased if the strength of the line depends on the continuum luminosity.

\subsection{The evolving galaxy population}\label{sec:evolvingresults}
Although we can measure the UV luminosity density with great accuracy, an estimate of the SFR Density depends sensitively on the dust extinction correction. Using the prescription from \citet{bouwens09}, we find that the SFR Density shows a significant increase by a factor of 4-5 between $z\sim 5$ and $z\sim 3$ . 

We find a strong increase in $\phi^{*}$ between $z\sim 5$ and
$z\sim 3$ by a factor of 2.5, which is a robust result. The characteristic magnitude, $M_{\rm{UV}}^{*}$, however is
not very well constrained by our data set. This is due to uncertainties in the 
redshift distributions of the source galaxies, as there are no spectroscopic data available.
Our data are consistent with a non-evolving $M_{\rm{UV}}^{*}$ ($\sim -20.9$) between $z\sim 5$ and
$z\sim 3$.
Our data do not show an evolution of the faint-end slope, and indicate $\alpha \sim -1.6$ in this epoch.

\section{Comparison with previous determinations}\label{sec:comparison}\label{sec:discussion}
\begin{figure}
\resizebox{\hsize}{!}{\includegraphics{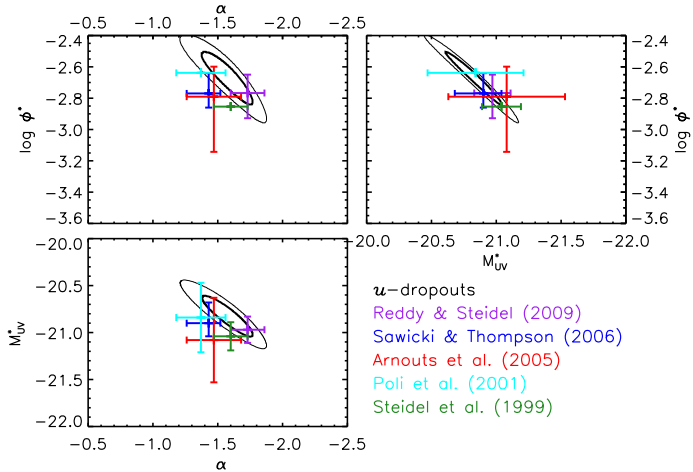}}
\caption{Comparison at $z\sim3$. The ellipses represent the 68\% and
  95\% confidence contours for different Schechter parameter
  combinations, based on our study. The error bars reflect the 68\%
  confidence limits on the results from previous studies. Our results
  agree within the 1-$\sigma$ level with most of the other
  determinations.}
\label{fig:testellipsume}
\end{figure}

\begin{figure}
\resizebox{\hsize}{!}{\includegraphics{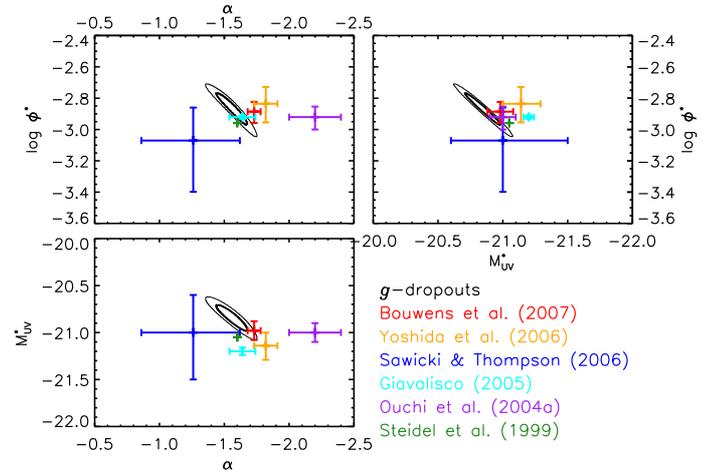}}
\caption{Similar as Fig. \ref{fig:testellipsume}, but now a comparison at $z\sim 4$.}
\label{fig:testellipsgme}
\end{figure}

\begin{figure}
\resizebox{\hsize}{!}{\includegraphics{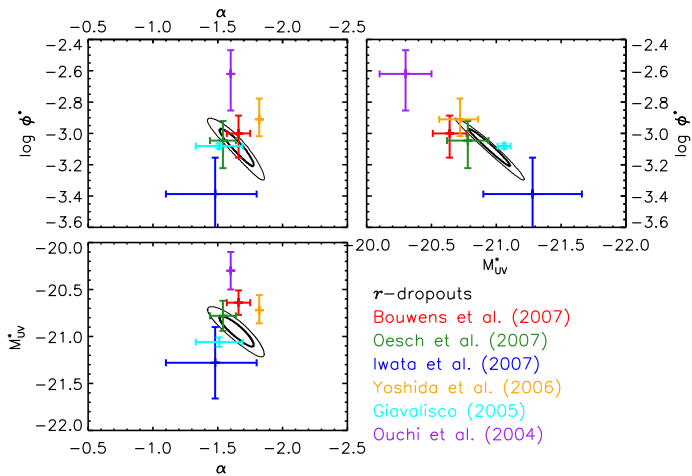}}
\caption{Similar as Fig. \ref{fig:testellipsume}, but now a comparison at $z\sim 5$.}
\label{fig:testellipsrme}
\end{figure}

Before we compare our results from each of the samples with Schechter parameters reported from
previous determinations in the literature, there are
a few things that should be noted. We will compare results at
redshifts of $around$ 3, 4 and 5. The LBGs are generally selected from different surveys and filter sets, 
and therefore intrinsically slightly different
galaxies may be selected, studied and compared. Also the rest-frame
wavelength at which the LF is estimated, varies. We compared the LFs
at 1600$\rm{\AA}$, while e.g. \citet{st2} measured the LF at
1700$\rm{\AA}$, and e.g. \citet{giavalisco04} did their analysis at
1500$\rm{\AA}$. Some of the studies described below make
use of identical or partially overlapping datasets. Our analysis, on
the contrary, is completely independent from previous
determinations, except for the dust extinction correction in our SFR density estimate. Furthermore the errors of several other studies could be
underestimated because only the Poisson noise component is taken into
account, as other noise components (e.g. cosmic variance) are
difficult to estimate with the typical small survey volumes of other
surveys. Also note that we compare our 2-dimensional error ellipses with 1-dimensional error bars that were attained after marginalizing over the two other Schechter parameters. This dilutes information on degeneracies between Schechter parameters. For these reasons the comparisons in this section will be
rather qualitative, and are meant to put our results into
context. Comparisons of the Schechter parameters are shown in Table
\ref{tab:refzall}, and in Figs. \ref{fig:testellipsume},
\ref{fig:testellipsgme} \& \ref{fig:testellipsrme}, and will be
discussed below. 

\begin{table*}[ht]
\caption{Our estimated Schechter parameters compared with values reported in the literature, for the three dropout samples.}
\label{tab:refzall}
\begin{center}
\centering
\begin{tabular}{cccc}
\hline\hline
Reference&$M_{UV}^{*}$& $\phi^{*}$ &$\alpha$\\
&&[$10^{-3}\, \rm{Mpc}^{-3}$]&\\
\multicolumn{3}{c}{$u$-dropouts}&\\
\hline
This work, Schechter&$-20.94^{+0.14}_{-0.13}$&$1.79^{+0.51}_{-0.38}$&$-1.65^{+0.12}_{-0.11}$\\
This work, Schechter+Magnification&$-20.84^{+0.15}_{-0.13}$&$2.11^{+0.63}_{-0.45}$&$-1.60^{+0.14}_{-0.11}$\\
\citet{reddy09}& $-20.97 \pm0.14$&$1.71\pm0.53$&$-1.73\pm0.13$\\
\citet{st2}&                   $-20.90^{+0.22}_{-0.14}$&$1.70^{+0.59}_{-0.32}$&$-1.43^{+0.17}_{-0.09}$\\
\citet{arnouts05}&       $-21.08\pm0.45$&$1.62\pm0.90$&$-1.47\pm0.21$\\
\citet{poli01}&              $-20.84\pm0.37$ &$2.3$&$-1.37\pm0.19$\\
\citet{steidel99}&         $-21.04 \pm 0.15$ &$1.4$&$-1.60 \pm 0.13$\\
\hline
\multicolumn{3}{c}{$g$-dropouts}&\\
\hline
This work&                       $-20.84 \pm 0.09$&$1.36^{+0.23}_{-0.20}$&$-1.56 \pm 0.08$\\
\citet{bouwens07}&    $-20.98 \pm 0.10$&$1.3 \pm 0.2$&$-1.73 \pm 0.05$\\
\citet{yoshida06}&      $-21.14^{+0.14}_{-0.15}$&$1.46^{+0.41}_{-0.35}$&$-1.82 \pm 0.09$\\
\citet{st2}&                   $-21.0^{+0.4}_{-0.5}$&$0.85^{+0.53}_{-0.45}$&$-1.26^{+0.40}_{-0.36}$\\
\citet{giavalisco05}&    $-21.20 \pm 0.04$&$1.20\pm0.03$&$-1.64\pm0.10$\\
\citet{ouchi04a}&           $-21.0 \pm 0.1$&$1.2 \pm0.2$&$-2.2 \pm 0.2$\\
\citet{steidel99}&          $-21.05$&$1.1$&$-1.6$ (fixed)\\          
\hline
\multicolumn{3}{c}{$r$-dropouts}&\\
\hline
This work&                       $-20.94^{+0.10}_{-0.11}$&$0.83^{+0.15}_{-0.14}$&$-1.65^{+0.09}_{-0.08}$\\
\citet{bouwens07}&    $-20.64\pm0.13$&$1.0\pm0.3$&$-1.66\pm0.09$\\
\citet{oesch07}&      $-20.78\pm0.16$&$0.9\pm0.3$&$-1.54\pm0.10$\\
\citet{iwata07}&                   $-21.28\pm0.38$&$0.41^{+0.29}_{-0.30}$&$-1.48^{+0.38}_{-0.32}$\\
\citet{yoshida06}&    $-20.72^{+0.16}_{-0.14}$&$1.23^{+0.44}_{-0.27}$&$-1.82$ (fixed)\\
\citet{giavalisco05}&       $-21.06\pm0.05$&$0.83\pm0.03$&$-1.51\pm0.18$\\
\citet{ouchi04a}&           $-20.3\pm0.2$&$2.4\pm1.0$&$-1.6$ (fixed)\\
\hline
\end{tabular}
\end{center}
\end{table*}

\subsection{Comparison at $z$=3}
We compare the results from our $u$-dropout sample with several
Schechter parameters reported in the literature. \citet{st2} estimated
the LF from the Keck deep fields, from $U_{n}GR$-selected star-forming
galaxies. Their survey area is about 60 times smaller than the
CFHTLS-Deep. The depth of their observations is slightly deeper than ours,
but due to our Eddington bias correction, we are able to probe the LFs
down to similar magnitudes. \citet{reddy09} estimate a LF at $z\sim3$,
from 31 spatially independent fields, having a total area of about a
quarter of ours. Their sample contains several thousands of
spectroscopic redshifts at $z$=2-3. \citet{arnouts05} mainly focussed
on the galaxy LF at lower redshifts ($0.2\le z \le1.2$) from
$GALEX$-data, but their redshift range was extended using 173 galaxies
at $z\sim3$ from the HDF sample. \citet{poli01} used the HDF-N, HDF-S
and NTT-DF samples to estimate the LF in the range $2.5\la z\la
3.5$. Their sample was therefore selected from a very small volume,
which makes their results susceptible to cosmic
variance. \citet{steidel99} pioneered this work, and estimated the UV
LF from 0.23 deg$^{2}$ of moderately deep data. Their study was
supported by a spectroscopic redshift sample. This data is included in
the study by \citet{reddy09}.

Our results agree within the $1-\sigma$ level with the results from
previous determinations at $z\sim3$. Note that the other data points lie in the direction of the elongated ellipse, and therefore in the direction of the degeneracy we find. 

\subsection{Comparison at $z$=4}
Our $g$-dropout sample is compared with the $z\sim4$ sample from
\citet{st2}, selected from the $GRI$ filter sets on the Keck
telescope. Their area is identical to the one from which the $z\sim3$
LF was estimated. \citet{steidel99} also estimated the $z\sim4$ LF
with a similar filter set as \citet{st2}, but did not probe deep
enough to be able to constrain the faint-end slope $\alpha$. This
parameter was set equal to the value found at $z=3$, namely
$\alpha=-1.6$. \citet{yoshida06} presented the LF for 3808
$BRi$'-selected LBGs, selected from the Subaru Deep Field
project. \citet{ouchi04a} selected a $z\sim 4$ LBG sample from Subaru
imaging, supported by a sample of 85 spectroscopically identified
objects. \citet{giavalisco04} used a $\sim$0.09 deg$^{2}$ sample from
the GOODS to estimate a LF for $B_{450}V_{606}z_{850}$-selected
LBGs. \citet{bouwens07} used the deep HST ACS fields, including the
HUDF and the GOODS, to select a sample of 4671 B-dropouts, from which
they estimated the UV LF to $M_{1600,AB}=-16.26$.

With this data set we have been able to measure the Schechter parameters for the $g$-dropout sample with very high statistical accuracy. Note however that several systematic uncertainties, which we comment upon in Sect. \ref{sec:robust}, are not included in our error ellipses.
Our results agree within the $1-\sigma$ level with many of the
$z\sim4$ results in the literature. However, there is still some
tension in measurements of the faint-end slope $\alpha$. The
characteristic magnitude, $\rm{M}_{\rm{UV}}^{*}$, we measure, is
slightly fainter than the values we found in the literature.

\subsection{Comparison at $z$=5}
At $z\sim5$, \citet{iwata07} reported the UV LF from a combination of
HDF and Subaru images, totalling a survey area about 1/9 of
ours. \citet{oesch07} based their study on approximately 100 LBGs from
very deep ACS and NICMOS imaging. \citet{yoshida06} also defined a
$z\sim 5$ sample from their observations, combining $Vi'z'$ and
$Ri'z'$ selected objects, as did
\citet{ouchi04a}. \citet{giavalisco04} selected 275
$V_{606}i_{775}z_{850}$ LBGs to estimate a $z\sim 5$ UV
LF. \citet{bouwens07} also measured a sample of 1416 V-dropouts from
their deep HST ACS sample, which resulted in an estimation of the UV
LF down to $M_{1600,AB}=-17.16$.

Similar to the Schechter parameters found for $z \sim 4$, there is a large discrepancy in the literature for the Schechter parameters at $z \sim 5$. The statistical uncertainties in the Schechter parameters is very small for our $r$-dropout sample. Note however that several systematic uncertainties are not included in these error ellipses, see Sect. \ref{sec:robust}. Our results agree reasonably well, within the $1-\sigma$ level, with many previous determinations at $z\sim 5$. 

\subsection{Comparison of the SFR density}\label{sec:SFRDcomparison}
\begin{figure}
\resizebox{\hsize}{!}{\includegraphics{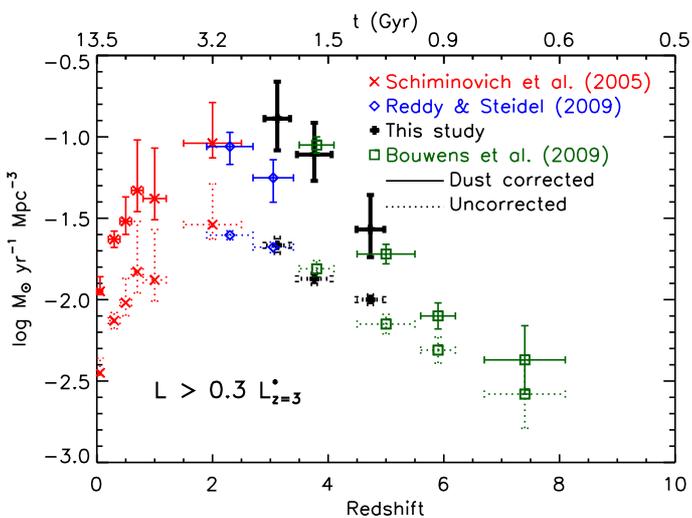}}
\caption{$\rho_{\rm{SFR}}$ as a function of redshift and cosmic
  time. Dotted: SFRD uncorrected for dust. Solid: the dust corrected SFRD, where we use a luminosity dependent 
  dust correction factor from \citet{bouwens09}. Note that we included both random and systematic errors in our dust correction, unlike most other studies.}
\label{fig:sfrd}
\end{figure}

In Fig.~\ref{fig:sfrd} we compare the SFR density values given in Table~\ref{tab:sfrd} 
to values reported by \citet{schiminovich05}, who made use of low-$z$ GALEX data,
\citet{reddy09} at intermediate $z$, and \citet{bouwens09} at high
$z$. The uncorrected SFRDs
are in good agreement with each other and show a smooth redshift
evolution. However, it is clear that the dust correction is the major uncertainty because of the age-dust degeneracy. We use the same dust correction as \citet{bouwens09} and also include systematic uncertainties in the error bars.

\section{Summary \& Conclusions}\label{sec:conclusions}
In this paper we use the CFHT Legacy Survey Deep fields to estimate
the UV Luminosity Functions of the largest $u$, $g$-, and $r$-dropouts
samples to date. As our samples are all extracted from the same
dataset this study is ideally suited to study a time evolution of the
luminosity function in the redshift regime $z=3-5$. Thanks to the
large volumes we probe with our 4 square degree survey, cosmic variance plays a negligible role in our analysis. We are now able to
study the bright end of the luminosity function with unprecedented
accuracy. Furthermore, given the depth of the stacked MegaCam images,
we probe the faint end of the luminosity function with comparable
precision as the deepest ground based surveys have done before. This
unique combination gives us the opportunity to not only estimate the
Schechter parameters for the different luminosity functions, but also
to study a possible deviation from this commonly used fit.

In the $u$-, and $g$-dropout samples of our survey we are able to measure the UV continuum slope directly from the data. This allows us to simulate sources that have the same distribution of UV slopes, which is important for an accurate estimate of the Schechter parameters.

We find the faint-end slope $\alpha$ to not evolve significantly in the redshift range we probe, and to have a value of around $-1.6$. This parameter however, is not very strongly constrained
by our ground based survey, as this parameter depends on some of the
assumptions made.

We do not find a significant evolution in $M_{\rm{UV}}^{*}$, and argue 
that this might be due
to insufficient knowledge of the redshift distribution of the source
galaxies. The conversion of apparent to absolute magnitudes depends
strongly on these distributions, and an uncertainty in the distance
modulus directly propagates into an equal uncertainty in
$M_{\rm{UV}}^{*}$. This parameter is therefore poorly constrained by
this study, until a more reliable redshift distribution is available.

We find a strong evolution in $\phi^{*}$, which we argue to be
significant. The normalisation of the LBG density, $\phi^{*}$,
increases by a factor of $\sim2.5$ from $z\approx5$ to $z\approx3$, an
increase that cannot be explained by any change in the assumptions
tested. We therefore conclude that the UV luminosity density is
increasing in the corresponding epoch, in a way that does not strongly
differ with magnitude.

The SFR Density does increase significantly, by a factor of $\sim 3$,
between $z\sim 5$ and $z\sim 4$. We find a smaller, but less significant 
increase between $z\sim 4$ and $z\sim 3$.

With our 4 square degree survey we probe densities that are at least
four times lower than any of the studies we compared our results
to. We find a substantial deviation from the Schechter function at the
bright end for the $u$-dropouts, where the LBG densities are very
low. We find that the deviation can be attributed to magnification effects that arise from inhomogeneities in the matter distribution between the LBGs and the observer. We fit an improved Schechter function that is corrected for magnification and find that the quality of the fit improves significantly. Intrinsically the distribution of luminosities does therefore not deviate significantly from a Schechter model. With this data set we have been the first to be able to measure a hint of this magnification imprint on a $z \sim 3$ LBG sample.

\begin{acknowledgements}
We thank Stefan Hilbert for supplying the magnification distribution for sources at our redshift of interest. We are grateful to Konrad Kuijken, Rychard Bouwens, Ludovic van Waerbeke and Marijn Franx for the interesting discussions we had and their suggestions to improve the
quality of this research. We thank Henk Hoekstra for the supply
  of powerful CPU's to perform the simulations with, and useful
  comments on the paper. We also thank the anonymous referee for a detailed report with nice suggestions for this paper.

We are grateful to the CFHTLS survey team for conducting the
observations and the TERAPIX team for developing software used in this
study. We acknowledge use of the Canadian Astronomy Data Centre
operated by the Dominion Astrophysical Observatory for the National
Research Council of Canada's Herzberg Institute of Astrophysics.  HH
is supported by the European DUEL RTN, project MRTN-CT-2006-036133.
TE is supported by the European DUEL RTN, the German Ministry
for Science and Education (BMBF) through the DESY project `GAVO III',
and the Deutsche Forschungsgemeinschaft through the projects SCHN 342/7 - 1 and ER 327/3 - 1 within the Priority Programme 1177.
\end{acknowledgements}

\bibliographystyle{aa} 
\bibliography{13812} 

\end{document}